\documentclass[journal=jacsat,manuscript=article]{achemso}

\usepackage[version=3]{mhchem} 
\usepackage{microtype}
\usepackage{siunitx}
\usepackage{booktabs}
\usepackage{amsmath}

\author{Frederico B. Sousa}
\affiliation{Departamento de F\'isica, Universidade Federal de São Carlos, São Carlos, São Paulo 13565-905, Brazil}
\email{fbsousa@ufscar.br}
\author{Boyang Zheng}
\affiliation{Department of Physics, The Pennsylvania State University, University Park, PA 16802, United States of America}
\alsoaffiliation{The Materials Research Institute, The Pennsylvania State University, University Park, PA 16802, United States of America}
\author{Elizabeth Grace Houser}
\affiliation{Department of Physics, The Pennsylvania State University, University Park, PA 16802, United States of America}
\alsoaffiliation{Center for 2-Dimensional and Layered Materials, The Pennsylvania State University, University Park, PA 16802, United States of America}
\author{Paulo E. Faria Junior}
\affiliation{Department of Electrical and Computer Engineering, University of Central Florida, Orlando, Florida 32816, USA}
\author{Zhuohang Yu}
\alsoaffiliation{Center for 2-Dimensional and Layered Materials, The Pennsylvania State University, University Park, PA 16802, United States of America}
\affiliation{Department of Materials Science and Engineering, The Pennsylvania State University, University Park, PA 16802, United States of America}
\author{Alessandra Ames}
\affiliation{Departamento de F\'isica, Universidade Federal de São Carlos, São Carlos, São Paulo 13565-905, Brazil}
\author{Gabriel A. D. Souza}
\affiliation{Departamento de F\'isica, Universidade Federal de São Carlos, São Carlos, São Paulo 13565-905, Brazil}
\author{Mingzu Liu}
\affiliation{Department of Physics, The Pennsylvania State University, University Park, PA 16802, United States of America}
\alsoaffiliation{Center for 2-Dimensional and Layered Materials, The Pennsylvania State University, University Park, PA 16802, United States of America}
\author{Gilmar Eugenio Marques}
\affiliation{Departamento de F\'isica, Universidade Federal de São Carlos, São Carlos, São Paulo 13565-905, Brazil}
\author{Leandro M. Malard}
\affiliation{Departamento de F\'isica, Universidade Federal de Minas Gerais, Belo Horizonte, Minas Gerais 30123-970, Brazil}
\author{Mauricio Terrones}
\affiliation{Center for 2-Dimensional and Layered Materials, The Pennsylvania State University, University Park, PA 16802, United States of America}
\alsoaffiliation{Department of Physics, The Pennsylvania State University, University Park, PA 16802, United States of America}
\alsoaffiliation{Department of Materials Science and Engineering, The Pennsylvania State University, University Park, PA 16802, United States of America}
\author{Vincent H. Crespi}
\affiliation{Department of Physics, The Pennsylvania State University, University Park, PA 16802, United States of America}
\alsoaffiliation{The Materials Research Institute, The Pennsylvania State University, University Park, PA 16802, United States of America}
\alsoaffiliation{Department of Materials Science and Engineering, The Pennsylvania State University, University Park, PA 16802, United States of America}
\alsoaffiliation{Center for 2-Dimensional and Layered Materials, The Pennsylvania State University, University Park, PA 16802, United States of America}
\author{Marcio D. Teodoro}
\affiliation{Departamento de F\'isica, Universidade Federal de São Carlos, São Carlos, São Paulo 13565-905, Brazil}
\email{mdaldin@ufscar.br}

\title{Vanadium doping induced valley asymmetries in \ce{WS2} monolayers}

\abbreviations{IR,NMR,UV}

\begin{document}







\newpage

\begin{abstract}

Transition metal dichalcogenide (TMD) monolayers offer an innovative platform for encoding and manipulating information through the valley degree of freedom. 
While unique valley-related physical phenomena have been reported so far, practical applications still require advanced control over the valley polarization efficiency and the valley Zeeman effect.
Recently, the introduction of spin-polarized metal atoms as substitutional defects was reported to break the time-reversal symmetry in TMD monolayers, consequently inducing a room-temperature ferromagnetic ordering and enhancing the valley-dependent optical responses.
Here, we report valley asymmetries for vanadium-doped \ce{WS2} monolayers.
With a given magnetic polarization, one valley exhibits larger Zeeman slope and degree of circular polarization than the other valley.
Additionally, the overall degree of circular polarization in the doped samples is approximately twice that of the pristine \ce{WS2} monolayer.
Density functional theory calculations in the doped structure show that different energy shifts in conduction band edges due to spin-dependent hybridization lead to different exciton energies between valleys, which is consistent with the experimental observations.
Our results pave the way for valleytronic technologies based on defect-engineered two-dimensional materials.


\end{abstract}

\section{Introduction}

Valleytronic technologies, i.e., based on the control and manipulation of the valley degree of freedom, gained a new paradigm with the emergence of two-dimensional (2D) materials~\cite{schaibley2016valleytronics}. 
Hexagonal 2D materials, such as graphene, exhibit inequivalent K and $-$K valleys with large momentum separation, offering a potential binary index for encoding and processing information~\cite{rycerz2007valley}. 
However, accessing valley-dependent phenomena in graphene is challenging due to its inversion symmetry~\cite{xiao2007valley,yao2008valley}.
In the case of semiconducting transition metal dichalcogenide (TMD) monolayers, the lack of an inversion center and the strong spin-orbit coupling result in unique valley physics~\cite{xiao2012coupled}.
For instance, TMD monolayers show a spin-valley locking effect and valley-dependent optical selection rules for circularly polarized light~\cite{xiao2012coupled,mak2012control}.
Moreover, under an external magnetic field, TMD monolayers experience opposite Zeeman shifts at K and $-$K valleys~\cite{Li2014,Aivazian2015,Srivastava2015,MacNeill2015}, leading to a breaking of valley degeneracy and enabling advanced opportunities for manipulating spin- and valley-polarized states. 
Because of the reduced dielectric screening and strong Coulomb interactions in TMD monolayers~\cite{wang2018colloquium}, the valley indices can be encoded by excitons even above room temperature.
However, valleytronics still faces fundamental challenges~\cite{xu2025valleytronics} regarding the short valley lifetimes, the valley polarization efficiency, and the high magnetic fields required for inducing a moderate valley Zeeman splitting in TMD monolayers.
Consequently, there is a significant technological interest in engineering alternatives that break the time-reversal symmetry in TMD monolayers, thereby enhancing control over the valley responses.

In this scenario, proximity effects induced by magnetic substrates are a promising approach for modulating valley phenomena, resulting, for example, in enhanced valley polarization, lifetime, and Zeeman splitting~\cite{norden2019giant,lyons2022giant,beer2024proximity,ali2025magnetic}, as well as valley asymmetries~\cite{serati2023charge}.
The introduction of spin-polarized metal atoms as substitutional defects in TMD monolayers is another alternative for breaking the time-reversal symmetry, which can lead to a room-temperature ferromagnetic ordering~\cite{ortiz2024transition}. 
This defect engineering was recently achieved through chemical vapor deposition (CVD)\cite{yun2020ferromagnetic,pham2020tunable,zhang2020monolayer}, a scalable growth method capable of furnishing large-area samples.
Particularly, an enhanced valley Zeeman splitting was observed for free excitons in Fe- and Co-doped MoS$_2$ monolayers~\cite{li2020enhanced,zhou2020synthesis} and for defect-bound excitons in a V-doped WSe$_2$ monolayer~\cite{sousa2024giant}, which also exhibits valley-dependent exciton and trion populations~\cite{nguyen2021spin}.
Moreover, a room-temperature valley degeneracy lifting was reported for V-doped MoS$_2$ monolayers even in the absence of an external magnetic field~\cite{sahoo2022enhanced,anil2025valley}.
These results point to the unique valley physics induced by the engineering of transition metal dopants.
Vanadium-doped \ce{WS2} monolayers also exhibit room-temperature magnetic ordering~\cite{zhang2020monolayer}, as well as strong modulation of excitonic responses~\cite{sousa2024effects,Mathela2025}, electron-phonon coupling~\cite{zou2023raman,sousa2026anisotropic}, electrical transport~\cite{gao2024electrical}, and nonlinear optical resonances~\cite{menescal2026resonance}.
However, valley-dependent phenomena in V-doped \ce{WS2} monolayers remain largely unexplored.

Here, we uncover pronounced valley asymmetries in \ce{WS2} monolayers induced by vanadium doping. 
Circularly polarized magneto-photoluminescence (magneto-PL) experiments reveal a threefold larger Zeeman shift in the $-$K valley compared with the K valley.
Moreover, the introduction of vanadium atoms results in an enhanced valley polarization, which is also asymmetric between K and $-$K valleys.
Density functional theory (DFT) calculations for the V-doped \ce{WS2} monolayer band structure indicate a distinct K and $-$K conduction band shift driven by the spin-dependent hybridization between the dopant $d_{z^2}$-orbital and the host \ce{WS2} conduction band edges, causing the different exciton energies between valleys at zero magnetic field.
Our results reveal unique valley-dependent optical responses, highlighting the potential of defect engineering in 2D materials for advanced valleytronic technologies with enhanced capabilities to control and manipulate information.

\section{Results and Discussion}

\begin{figure}[htbp]
\centering
\includegraphics[width=1.0\textwidth]{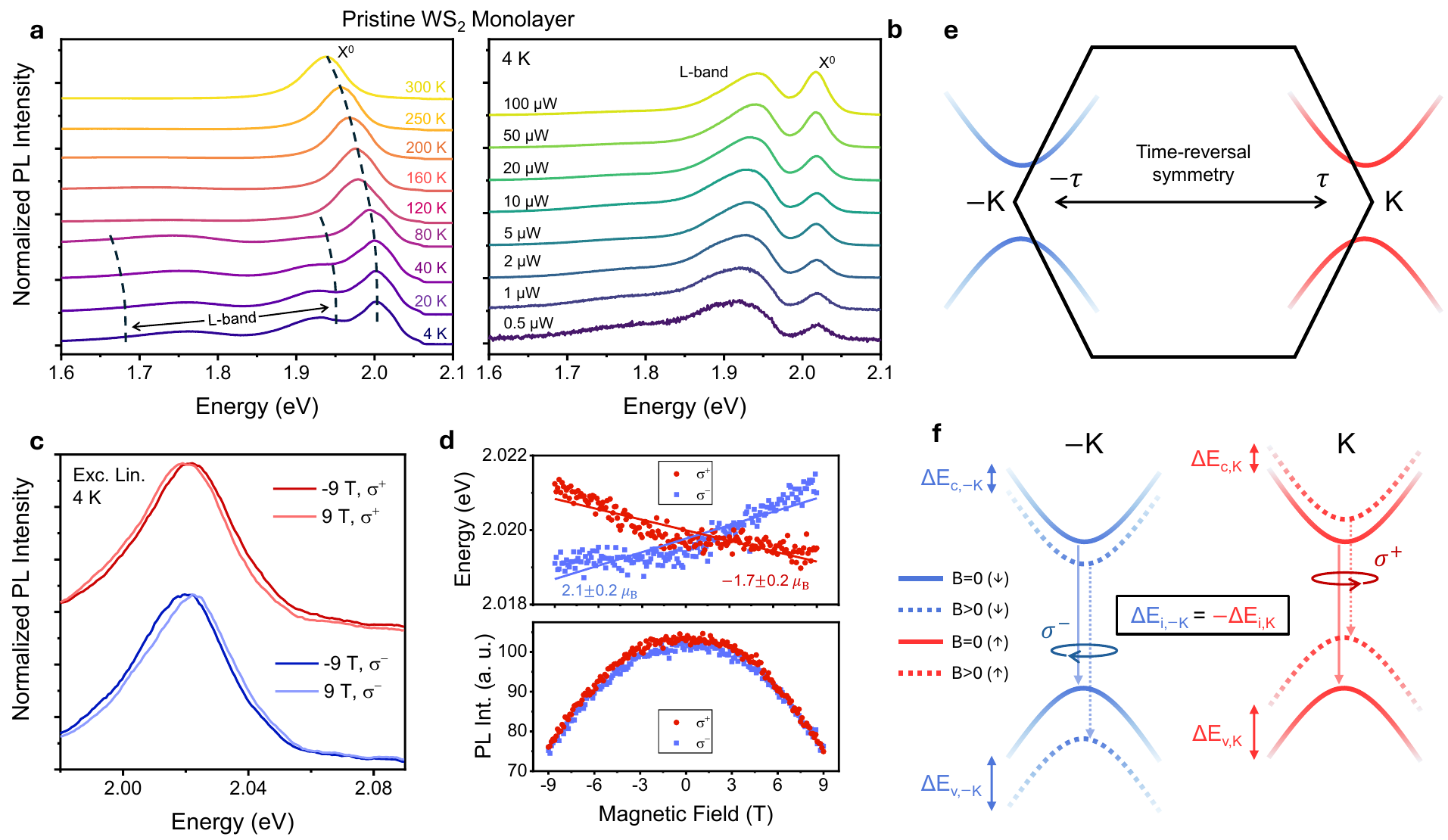} 
\caption{{\small {\bf a} Temperature- and {\bf b} power-dependent normalized PL spectra for a pristine \ce{WS2} monolayer displaying X$_0$ and L-band emissions. 
{\bf c} Normalized PL spectra acquired at \qty{9}{T} and \qty{-9}{T}, \qty{4}{K}, linearly polarized excitation, and circular polarization detection for the X$_0$ emission in the pristine \ce{WS2} monolayer.
Red (blue) line spectra denote $\sigma^+$ ($\sigma^-$) detection.
{\bf d} Magnetic-field-dependent X$_0$ energy (top panel) and intensity (bottom panel) graphs for both $\sigma^+$ and $\sigma^-$ signals.
Similar Zeeman slopes of $\alpha_K=$ \qty{-1.7\pm0.2}{\mu_{\textnormal{B}}} and $\alpha_{-K}=$ \qty{2.1\pm0.2}{\mu_{\textnormal{B}}} are obtained for K and $-$K valleys, respectively.
{\bf e} Representation of the first Brillouin zone of a pristine \ce{WS2} monolayer with K and $-$K valleys connected through time-reversal symmetry.
{\bf f} Valley Zeeman effect sketch for a pristine \ce{WS2} monolayer, highlighting the opposite band shift but with the same magnitude at K and $-$K valleys under an external magnetic field.
}}
\label{fig1}
\end{figure}

The pristine and V-doped \ce{WS2} monolayers studied here were synthesized through CVD, as described in the Methods Section.
Scanning transmission electron microscopy (STEM) experiments were employed in a previous work~\cite{sousa2024effects} for these samples, revealing \ce{WS2} monolayers with average vanadium concentrations of \qty{\sim0.4}{\percent} and \qty{\sim2.0}{\percent}.
Figures~\ref{fig1}a,b show temperature- and power-dependent photoluminescence (PL) spectra for the pristine \ce{WS2} monolayer.
As expected, the free exciton (X$_0$) emission displays a linear power dependence (see Supporting Figure~S1) and a redshift under increasing temperature. 
A lower-energy emission, commonly referred as the L-band, exhibits a sublinear power dependence and a finite signal only at low temperatures (see Supporting Figure~S1), a fingerprint of localized defect-related peaks~\cite{sousa2025optical}.
Even for TMD monolayers without the intentional introduction of defects, distinct impurities might be adsorbed due to ambient exposure, which are the origin of the observed L-band~\cite{rogers2018laser,venanzi2019exciton,mitterreiter2021role}.
Since a magneto-optical characterization of the L-band has been recently reported~\cite{sousa2025strong}, we focus on the X$_0$ emission of the pristine monolayer.

Magneto-PL experiments were performed at \qty{4}{K}, linearly polarized excitation (\qty{2.33}{eV}, \qty{100}{{\mu}W}), and \qty{0.1}{T} steps from \qty{9}{T} to \qty{-9}{T} in a Faraday geometry.
Circularly polarized detection was employed to selectively access emissions from K ($\sigma^+$) and $-$K ($\sigma^-$) valleys.
Figure~\ref{fig1}c shows PL spectra (normalized by the maximum intensity) obtained at \qty{9}{T} and \qty{-9}{T} for the pristine \ce{WS2} monolayer, revealing a similar Zeeman shift for both valleys.
To better analyze this response, a Gaussian function was fitted to each magneto-PL spectrum (see Supporting Figure~S2), and the extracted energy and integrated intensity values for the X$_0$ emission are presented in Figure~\ref{fig1}d.
Considering the exciton Zeeman shift in $i=\{K,-K\}$ valleys as $\Delta E_{i} = \alpha_{i} B$, where $B$ is the applied magnetic field, we obtained Zeeman slopes of $\alpha_K=$ \qty{-1.7\pm0.2}{\mu_{\textnormal{B}}} and $\alpha_{-K}=$ \qty{2.1\pm0.2}{\mu_{\textnormal{B}}}. 
Expressed in units of the Bohr magneton constant ($\mu_{\textnormal{B}}$), the two slopes have similar magnitudes within the experimental uncertainty.
Additionally, they yield an effective exciton g-factor of $g = -3.8 \pm 0.4$, in accordance with previous reports~\cite{Li2014,Aivazian2015,Srivastava2015,MacNeill2015}.
The magnetic field modulation of the PL intensity is also similar for both $\sigma^+$ and $\sigma^-$ detections.
Since K and $-$K valleys are connected through time-reversal symmetry in pristine TMD monolayers (see Figure~\ref{fig1}e), their opposite spins lead to opposite Zeeman shifts with the same magnitude~\cite{Li2014,Aivazian2015,Srivastava2015,MacNeill2015} (see Figure~\ref{fig1}f), as observed in our experiment.

\begin{figure}[htbp]
\centering
\includegraphics[width=1.0\textwidth]{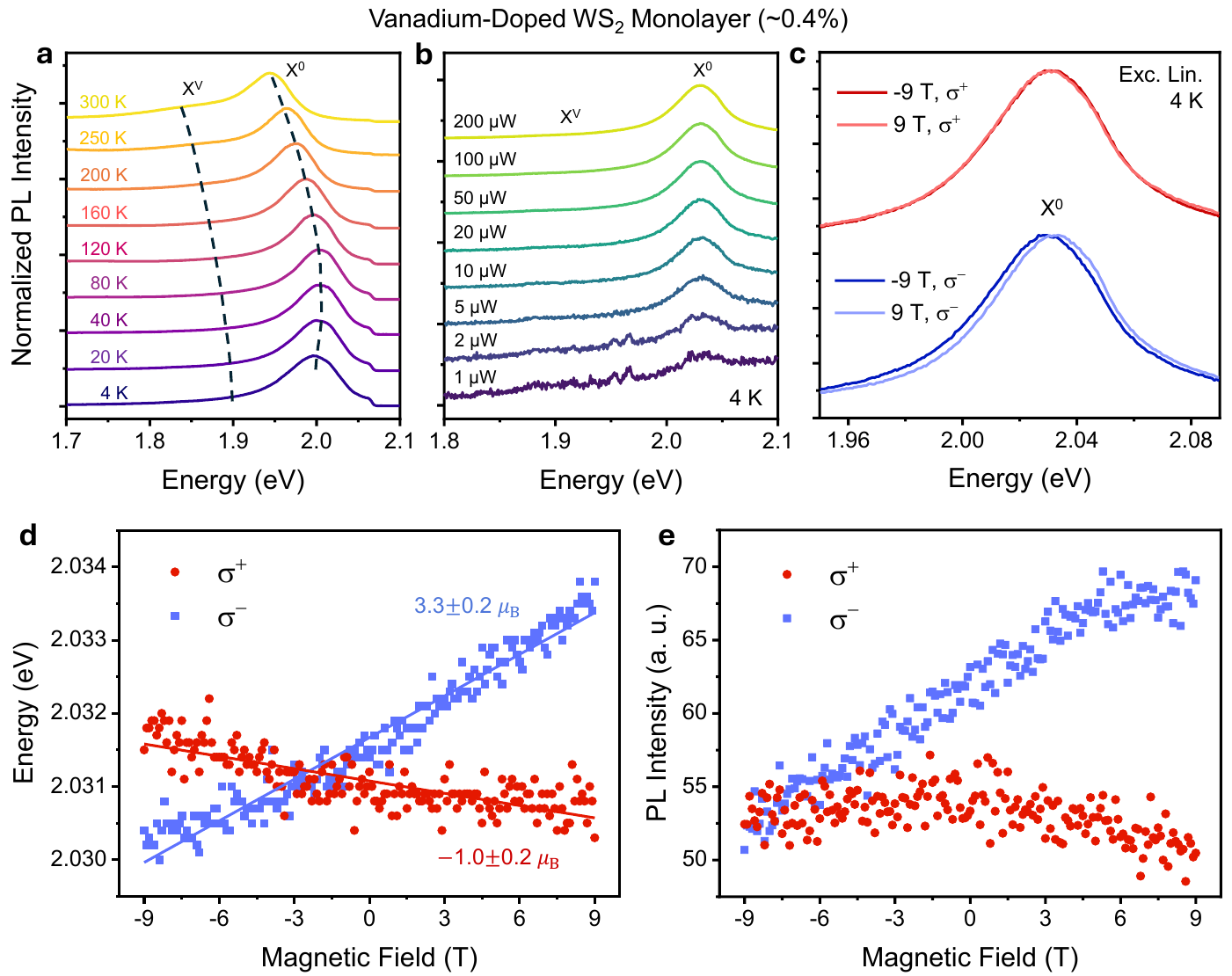} 
\caption{{\small {\bf a} Temperature- and {\bf b} power-dependent normalized PL spectra for a \qty{\sim0.4}{\percent} V-doped \ce{WS2} monolayer displaying X$_0$ and X$_V$ emissions.
{\bf c} Normalized PL spectra acquired at \qty{9}{T} and \qty{-9}{T}, \qty{4}{K}, with linearly polarized excitation, and circular polarization detection for the X$_0$ emission.
Red (blue) line spectra denote $\sigma^+$ ($\sigma^-$) detection.
{\bf d} Magnetic-field-dependent X$_0$ energy graph for both $\sigma^+$ and $\sigma^-$ signals.
Valley-dependent Zeeman slopes of $\alpha_K=$ \qty{-1.0\pm0.2}{\mu_{\textnormal{B}}} and $\alpha_{-K}=$ \qty{3.3\pm0.2}{\mu_{\textnormal{B}}} are obtained for K and $-$K valleys, respectively.
{\bf e} Magnetic-field-dependent X$_0$ integrated intensity graph for both $\sigma^+$ and $\sigma^-$ signals corroborating the valley asymmetric response.
}}
\label{fig2}
\end{figure}

To investigate the impact of vanadium doping and the consequent time-reversal symmetry breaking in the valley Zeeman effect of \ce{WS2} monolayers, we performed similar PL and magneto-PL experiments for the doped samples. 
Figures~\ref{fig2}a,b show temperature- and power-dependent PL spectra for the \qty{\sim0.4}{\percent} V-doped \ce{WS2} monolayer.
Following a recent report~\cite{sousa2024effects}, we identify X$_0$ and vanadium-related defect-bound exciton (X$_V$) emissions. 
In contrast to the L-band, the X$_V$ peak exhibits a finite signal at room temperature (see Supporting Figure~S3).
However, the weak X$_V$ intensity at low temperatures precludes a reliable analysis of its spectral features. 
We therefore focus our discussion on the X$_0$ peak of the doped sample as well.

Figure~\ref{fig2}c shows PL spectra (normalized by the maximum intensity) for the X$_0$ emission in the \qty{\sim0.4}{\percent} V-doped \ce{WS2} monolayer acquired at \qty{9}{T} and \qty{-9}{T}, \qty{4}{K}, with linearly polarized excitation, for $\sigma^+$ and $\sigma^-$ polarizations in detection.
Interestingly, the Zeeman shift at the $-$K valley is larger than that at the K valley.
This asymmetry is better noted in Figure~\ref{fig2}d, where the X$_0$ energies (extracted by fitting the spectra with a Gaussian function, see Supporting Figure~S2) are plotted for all measured magnetic fields. 
Through a linear regression, we obtained valley-dependent Zeeman slopes of $\alpha_K=$ \qty{-1.0\pm0.2}{\mu_{\textnormal{B}}} and $\alpha_{-K}=$ \qty{3.3\pm0.2}{\mu_{\textnormal{B}}} for K and $-$K valleys, respectively.
The modulation of the X$_0$ integrated intensity by the external magnetic field is also valley asymmetric, as shown in Figure~\ref{fig2}e, with the $-$K valley presenting a greater overall intensity.
Supporting Figure~S4 shows similar magneto-PL experiments obtained for different flakes, confirming the reproducibility of the asymmetric valley response.
In these measurements, no qualitative differences were observed between negative-to-positive and positive-to-negative magnetic field sweep directions, indicating that the observed valley asymmetry is not affected by the magnetic-field history and suggesting a coercive field exceeding the maximum applied field of \qty{9}{T}.
This valley-dependent phenomenon is also observed for the \qty{\sim2.0}{\percent} V-doped \ce{WS2} monolayer (see Supporting Figure~S5), with Zeeman slopes of $\alpha_K=$ \qty{-0.5\pm0.2}{\mu_{\textnormal{B}}} and $\alpha_{-K}=$ \qty{2.6\pm0.2} for K and $-$K valleys, respectively.

\begin{figure}[htbp]
    \centering
    \includegraphics[width=1.0\textwidth]{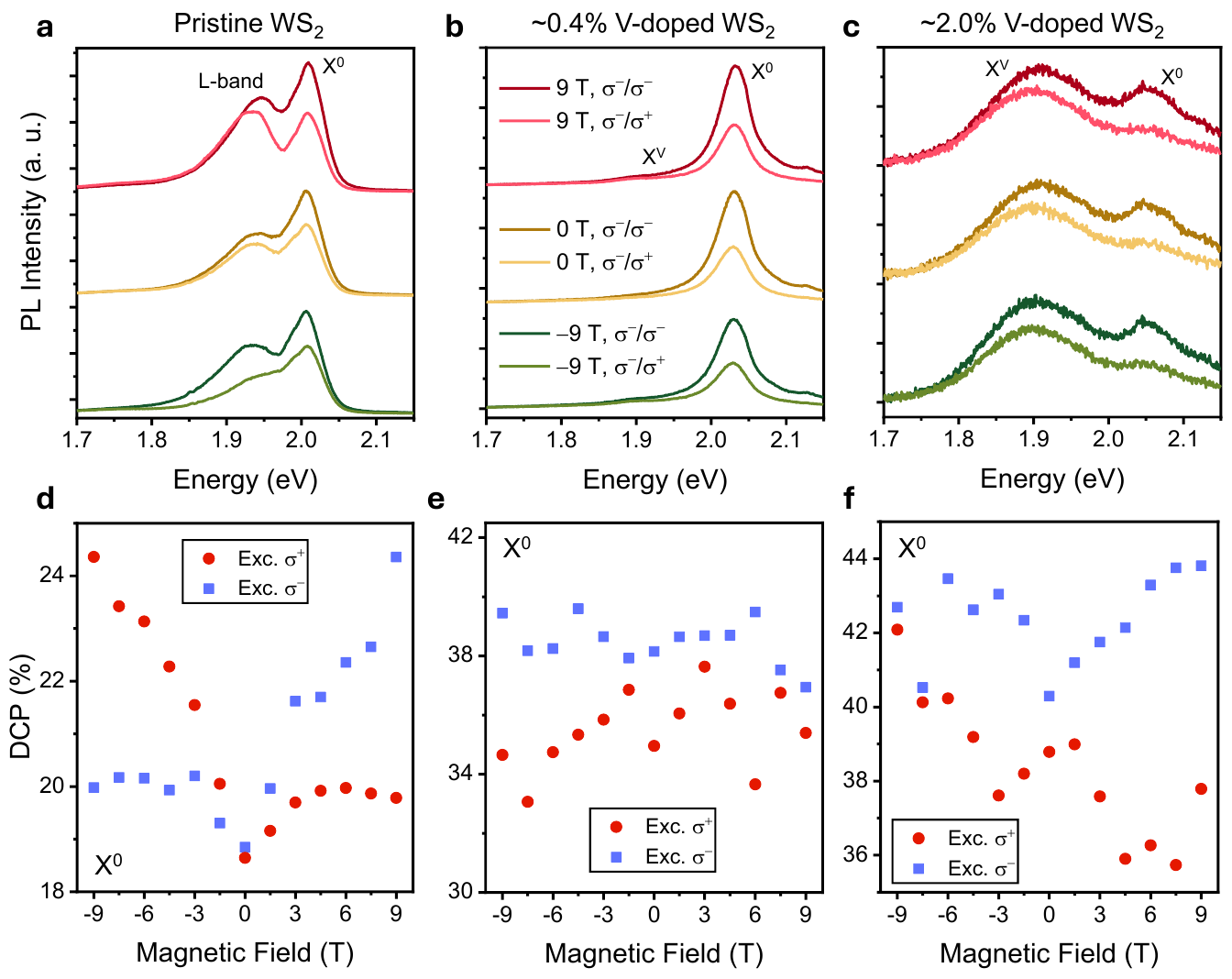} 
    \caption{{\small {\bf a-c} PL spectra acquired at \qty{4}{K}, with magnetic field at \qtylist{9;0;-9}{T} for the pristine ({\bf a}), \qty{\sim0.4}{\percent} ({\bf b}), and \qty{\sim2.0}{\percent} ({\bf c}) V-doped \ce{WS2} monolayers. 
    The spectra were obtained with $\sigma^-$ excitation and both $\sigma^-$ and $\sigma^+$ detections (indicated by exc./det.\ as $\sigma^-/\sigma^-$ and $\sigma^-/\sigma^+$).
    {\bf d-f} Magnetic-field-dependent DCP graphs for the pristine ({\bf d}), \qty{\sim0.4}{\percent} ({\bf e}), and \qty{\sim2.0}{\percent} ({\bf f}) V-doped \ce{WS2} monolayers.
    The DCPs are shown for both $\sigma^+$ ($\mathrm{DCP_{K}}$) and $\sigma^-$ ($\mathrm{DCP_{-K}}$) excitations, revealing the symmetric and asymmetric responses for pristine and doped monolayers, respectively.
    }}
    \label{fig3}
\end{figure}

To further gain insights into the valley physics in V-doped \ce{WS2} monolayers, we acquired PL spectra with circularly polarized excitation and detection at \qty{4}{K}, and swept the applied field from \qty{9}{T} to \qty{-9}{T} with a \qty{1.5}{T} step size, as shown in Fig.~\ref{fig3}.
The circularly polarized excitation (for example, $\sigma^-$) induces an initial population at the corresponding valley ($-$K), and the valley depolarization can be studied by comparing the relative PL intensity between $\sigma^-$ and $\sigma^+$ signals~\cite{mak2012control}.
Therefore, we can quantify the valley polarization by calculating the magnetic-field-dependent degree of circular polarization (DCP), given by
\begin{align}
    \mathrm{DCP_{K}(\%)} &= 100\frac{I_{\sigma^+/\sigma^+}-I_{\sigma^+/\sigma^-}}{I_{\sigma^+/\sigma^+}+I_{\sigma^+/\sigma^-}}, \\
    \mathrm{DCP_{-K}(\%)} &= 100\frac{I_{\sigma^-/\sigma^-}-I_{\sigma^-/\sigma^+}}{I_{\sigma^-/\sigma^-}+I_{\sigma^-/\sigma^+}},
\end{align}
in which $I_{\sigma^i/\sigma^j}$ is the PL integrated intensity obtained from spectral fitting with $\sigma^i$ excitation and $\sigma^j$ detection.
Figures~\ref{fig3}a--c show PL spectra acquired at \qty{9}{T}, \qty{0}{T}, and \qty{-9}{T} for the pristine, \qty{\sim0.4}{\percent}, and \qty{\sim2.0}{\percent} V-doped \ce{WS2} monolayers using a $\sigma^-$ excitation and detecting at both $\sigma^-$ and $\sigma^+$.
Using the fitting procedure shown in Supporting Figure~S2, we extracted the integrated intensities from the obtained spectra to determine the X$_0$ $\mathrm{DCP_{K}}$ and $\mathrm{DCP_{-K}}$ for pristine and doped samples, which are presented in Figures~\ref{fig3}d--f.
The pristine sample exhibits similar DCP values of $\sim$19$\%$ at \qty{0}{T} for both K and $-$K valleys, and increases up to $\sim$24$\%$ at \qty{9}{T} (\qty{-9}{T}) for $\sigma^-$ ($\sigma^+$) excitation.
Although the valley polarization at K and $-$K points varies distinctly with magnetic field, they are symmetric with respect to \qty{0}{T}, similarly to the valley Zeeman splitting shown in Figure~\ref{fig1}d.
In contrast, the V-doped \ce{WS2} monolayers present a valley polarization asymmetry where the $-$K valley shows a larger polarization than the K valley over the whole range of applied field.
Moreover, vanadium doping also induces an overall enhancement of the valley polarization compared to the pristine sample, with a $\mathrm{DCP_{-K}}/\mathrm{DCP_{K}}$ of $\qty{\sim39}{\percent}/\qty{\sim36}{\percent}$ and $\qty{\sim43}{\percent}/\qty{\sim39}{\percent}$ for the \qty{\sim0.4}{\percent} and \qty{\sim2.0}{\percent} V-doped \ce{WS2} monolayers, respectively, as compared to $\qty{\sim19}{\percent}/\qty{\sim19}{\percent}$ for the pristine at \qty{0}{T}.
These experimental findings unveil the unique valley physics of vanadium-doped \ce{WS2} monolayers with a promising appeal to enhance valley polarization efficiency and offer routes to selectively manipulate valleys in 2D semiconductors.

\begin{figure}[htbp]
\centering
\includegraphics[width=1.0\textwidth]{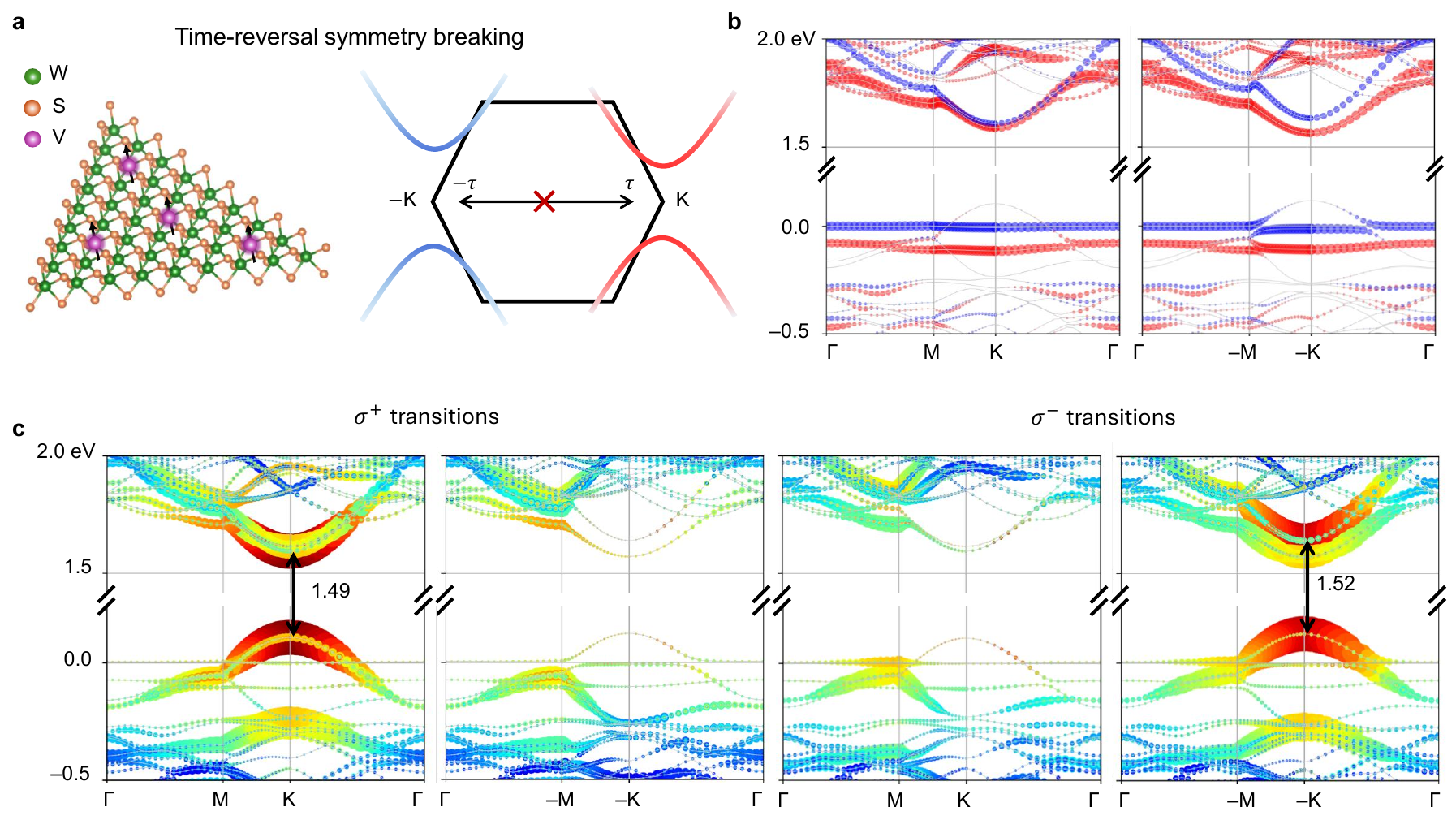} 
\caption{{\small {\bf a} Schematics of a vanadium-doped \ce{WS2} monolayer and the consequent time-reversal symmetry breaking between K and $-$K valleys.
{\bf b} The band structure of a \qty{\sim4}{\percent} V-doped \ce{WS2} monolayer.
    The areal size of the circles is proportional to the projection to the V $dz^2$ orbital, with the red (blue) color denoting spin-up (-down) states.
{\bf c} Circularly polarized optical transitions calculated for the V-doped \ce{WS2} monolayer.
    The areal size for pairs of circles (one in the conduction band, and one in the valence band for the same $k$) is proportional to the magnitude of the transition dipole moment, with the color denoting the energy of the transition.
}}
\label{fig4}
\end{figure}

As previously discussed, the K and $-$K valleys in pristine \ce{WS2} monolayers are related by time-reversal symmetry and therefore exhibit symmetric responses under opposite magnetic fields. 
Incorporating magnetic vanadium atoms into the \ce{WS2} lattice with an out-of-plane magnetic moment breaks symmetry relations between these two valleys (Figure~\ref{fig4}a, and Supporting Note~6), making it possible to have the pronounced asymmetries measured experimentally.
Distinct valley-dependent phenomena have recently been reported in V-doped MoS$_2$ monolayers, where asymmetric intervalley dynamics were attributed to the lifting of this valley degeneracy~\cite{sahoo2022enhanced,anil2025valley,kamath2026spin}.
To gain deeper insights into the specific mechanisms responsible for the valley asymmetry in V-doped \ce{WS2}, we performed DFT calculations of its electronic band structure.
In our DFT calculations, after the introduction of the dopant, we observe different energy shifts for conduction band edges, as shown in Figure~\ref{fig4}b, where the supercell has a total magnetic moment of \qty{0.52}{\mu_{\mathrm{B}}} along the $z$-direction.
In the K valley, the hybridization between the spin-up vanadium $d_{z^2}$-orbital and the host \ce{WS2} drops the spin-up conduction band edge lower in energy, to the extent that it is even below the spin-down conduction band edge in our calculation.
On the other hand, the $-$K valley spin-down conduction band edge is not as much affected by the vanadium dopant.
These two conduction band edges are responsible for the $\sigma^+$ emission at K and the $\sigma^-$ emission at $-$K valleys, respectively (Figure~\ref{fig4}c), and their distinct energy shifts lead to a smaller exciton energy for the $\sigma^+$ emission compared to the $\sigma^-$ emission at zero external field, consistent with the experimental observation.

For the magneto-optical response, the valley Zeeman shift of the exciton energy is~\cite{Aivazian2015}
\begin{equation}
    \Delta E = \left(-2 + \frac{m_0}{m_{\textnormal{c}}} - \frac{m_0}{m_\textnormal{v}}\right)\tau \mu_B B \label{eq:Zeeman-shift}
\end{equation}
where $\tau$ is the valley index ($\tau = +1$ for K and $\tau = -1$ for $-$K), $m_0$ is the free electron mass, $m_\textnormal{c}$ and $m_\textnormal{v}$ are the effective masses of the conduction and valence bands, respectively, $\mu_B$ is the Bohr magneton, and $B$ is the magnetic field.
By calculating the relative change of the effective masses for the V-doped system compared to the pristine \ce{WS2} monolayer (see Methods Section), the Zeeman slope is estimated to be $\qty{-1.94}{\mu_{\textnormal{B}}}$ for the K valley and $\qty{1.90}{\mu_{\textnormal{B}}}$ for the $-$K valley.
These values indicate a valley asymmetry of the Zeeman slopes, although they significantly differ from the ones observed in the experiments (Fig.~\ref{fig2}d).
Notably, Eq.~\ref{eq:Zeeman-shift} shows that the Zeeman slope deviates from $\qty{2}{\mu_{\textnormal{B}}}$ by the effective mass difference between the conduction and valence bands.
For pristine \ce{WS2}, the Zeeman slope from Fig.~\ref{fig1} shows that the effective mass difference would contribute to the order of \qty{0.1}{\mu_{\textnormal{B}}}.
Although we have not included any band gap correction in our calculation, we do not expect that a low (\qty{\sim1}{\percent}) doping level would significantly change this effective mass difference such that the overall Zeeman slope could deviate from the $\qty{2}{\mu_{\textnormal{B}}}$ value to the order of \qty{1}{\mu_{\textnormal{B}}}.
Therefore, while our calculations qualitatively corroborate the V-doping-induced valley asymmetry, the quantitative discrepancies require further investigation to fully understand the valley Zeeman effect of this system.

\section{Conclusion}
 
In summary, we demonstrate robust valley asymmetries in vanadium-doped \ce{WS2} monolayers through circular polarization magneto-PL experiments.
While pristine \ce{WS2} monolayers show valley Zeeman shifts with the same magnitude between the valleys, the doped samples exhibit a threefold larger Zeeman slope at the $-$K valley compared to the K valley.
An asymmetric response was also noted for the valley polarization of the V-doped monolayers measured by the DCP, where the $-$K valley possesses larger values than the K valley across the whole range of magnetic field tested.
In addition, vanadium doping also induces an enhancement of the valley polarization, with DCPs measured up to \qty{\sim39}{\percent} and \qty{\sim43}{\percent} for \qty{\sim0.4}{\percent} and \qty{\sim2.0}{\percent} V-doped \ce{WS2} monolayers, respectively, against a DCP of \qty{\sim19}{\percent} for the pristine sample at \qty{0}{T}.
DFT calculations reveal a spin-dependent hybridization between the vanadium dopant and the host \ce{WS2} bands, which accounts for the valley asymmetry at zero field.
Our results uncover a novel alternative for modulating valley phenomena through the defect engineering of 2D TMDs, highlighting the potential of vanadium-doped \ce{WS2} monolayers for advanced valleytronic technologies.

\section{Methods}

\subsection{Sample Preparation}

Monolayer V-doped \ce{WS2} flakes were synthesized using a liquid-assisted chemical vapor deposition (CVD) method, following the procedure reported previously~\cite{zhang2020monolayer}. The W precursor solution was prepared by dissolving ammonium metatungstate hydrate ((NH$_4$)$_6$H$_2$W$_{12}$O$_{40}$·xH$_2$O) and sodium cholate hydrate (C$_{24}$H$_{39}$NaO$_5$·xH$_2$O) in deionized water. Separately, vanadyl sulfate (VO[SO$_4$]) was dissolved in deionized water to obtain the V precursor solution. The W and V precursor solutions were then mixed and spin-coated onto SiO$_2$/Si substrates. The prepared substrates were transferred into a quartz tube furnace, while sulfur powder was placed upstream to serve as the sulfur source. Growth was carried out at 825~°C for 15~min under a continuous flow of ultrahigh-purity Ar carrier gas. Following the growth process, the furnace was allowed to cool naturally to room temperature while maintaining the Ar atmosphere. The vanadium concentration in the samples was tuned by varying the volume ratio of the W and V precursor solutions.

\subsection{Spectroscopy Measurements}

The PL spectroscopy experiments were carried out using a 532~nm (2.33~eV) diode laser with an excitation power of 100~$\mu$W (except where explicitly stated otherwise) in confocal microscopes coupled to cryostats.
For the temperature-dependent PL, the sample was placed in a cryostat (Attocube --- Attodry 800) capable of varying the temperature from \qty{4}{K} to \qty{300}{K}. 
For the magneto-PL measurements, a magneto-cryostat (Attocube --- Attodry 1000) that applies magnetic fields up to \qty{9}{T} at \qty{4}{K} was used.
The applied magnetic fields were perpendicular to the sample plane (Faraday geometry).
The linear and circular polarizations at both excitation and detection were controlled through a set of half-wave plates, quarter-wave plates, and polarizers.
The PL signal was detected in a spectrometer equipped with a sensitive CCD camera (Andor --- Shamrock-Idus).

\subsection{Calculations}

Spin-orbit-coupled density functional theory (DFT) calculations were performed using the Vienna Ab initio Simulation Package (VASP)~\cite{kresseInitioMolecularDynamics1993,kresseEfficientIterativeSchemes1996,kresseEfficiencyAbinitioTotal1996,kresseUltrasoftPseudopotentialsProjector1999} with the projector augmented wave (PAW) method~\cite{blochlProjectorAugmentedwaveMethod1994}, and the exchange-correlation potential was described by the Perdew-Burke-Ernzerhof (PBE) functional~\cite{perdewGeneralizedGradientApproximation1996}.
\numproduct{5x5} \ce{WS2} supercell with one V dopant were used with the unit cell parameters fixed to the values of pristine \ce{WS2} monolayers ($a=\qty{3.188}{\angstrom}$, $c=\qty{15}{\angstrom}$).
The initial spin polarization \qty{1}{\mu_{\textnormal{B}}} is given to the vanadium atom along the $z$-axis, and the total magnetic moment of the system relaxes to \qty{0.39}{\mu_{\textnormal{B}}} after the self-consistent field (SCF) calculation.
The cutoff energy was set to \qty{500}{eV} and the Brillouin zone was sampled with a \numproduct{5x5x1} Monkhorst-Pack grid~\cite{monkhorstSpecialPointsBrillouinzone1976}.
The SCF calculation was performed with a convergence criterion of \qty{1e-7}{eV} for the total energy, and the atomic positions were relaxed until the forces were smaller than \qty{0.01}{eV/\angstrom}.
The optical transitions were calculated with home-made codes modified from the code \texttt{VaspBandUnfolding}~\cite{zhengVaspBandUnfolding2023}.

To get the effective masses at conduction band edges, one may need to accurately calculate the band energies using either the GW method or a hybrid functional, which is computationally very expensive to do with a supercell system with spin-orbit coupling.
In this work, we instead notice that the Zeeman slope can be rewritten as 
\begin{align}
    \textnormal{Zeeman slope} &= \left[-2+C\left(E''_{\textnormal{c}}(k)+E''_{\textnormal{c}}(v)\right)\right] \tau \mu_{\textnormal{B}}
\end{align}
where $C$ is some constant containing all relevant physical constants and corrections.
Assuming the $C$ value is the same for both the pristine and the doped supercell under the same level of computational accuracy, we can estimate the Zeeman slope for our doped model with the known value \qty{1.9}{\mu_{\textnormal{B}}} (1.7 and 2.1 averaged to 1.9 in Fig.~\ref{fig1}) for pristine \ce{WS2}.


\begin{acknowledgement}

F.B.S., G.A.D.S., A.A., G.E.M., and M.D.T.\ acknowledge support from FAPESP (2022/10340-2, 2024/15969-1, 2025/02343-0).
F.B.S., G.A.D.S., A.A., G.E.M., L.M.M., and M.D.T.\ acknowledge financial support from the Brazilian agencies CNPq and CAPES.
L.M.M.\ thanks FAPEMIG, FINEP, the National Institute of Science and Technology (INCT) in Carbon Nanomaterials, and the Rede Mineira de Materiais 2D (FAPEMIG) for funding.
B.Z.\ and V.H.C.\ thank the support from Two-Dimensional Crystal Consortium-Materials Innovation Platform (2DCC-MIP) under NSF cooperative agreement no.\ DMR-2039351.
Z.Y., M.L., E.G.H., and M.T.\ acknowledge financial support from AFOSR (FA9550-23-1-0447).

\end{acknowledgement}





\providecommand{\latin}[1]{#1}
\makeatletter
\providecommand{\doi}
  {\begingroup\let\do\@makeother\dospecials
  \catcode`\{=1 \catcode`\}=2 \doi@aux}
\providecommand{\doi@aux}[1]{\endgroup\texttt{#1}}
\makeatother
\providecommand*\mcitethebibliography{\thebibliography}
\csname @ifundefined\endcsname{endmcitethebibliography}  {\let\endmcitethebibliography\endthebibliography}{}

\clearpage

\setcounter{figure}{0}
\renewcommand{\figurename}{\text{Supplementary Figure}}
\renewcommand{\thefigure}{\text{S\arabic{figure}}}

\renewcommand{\thetable}{\text{S\arabic{table}}}

\setcounter{equation}{0}
\renewcommand{\theequation}{S\arabic{equation}}



\title{}
{\noindent\large{\textbf {%
\centering{%
Supporting Information for \\
Vanadium doping induced valley asymmetries in WS$_2$ monolayers
}}}
}

\paragraph*{}

\centering{Frederico B. Sousa$^{1}$*, Boyang Zheng$^{2,3}$, Elizabeth Grace Houser$^{2,4}$, Paulo E. Faria Junior$^{5}$, Zhuohang Yu$^{4,6}$, Alessandra Ames$^{1}$, Gabriel A. D. Souza$^{1}$, Mingzu Liu$^{2,4}$, Gilmar Eugenio Marques$^{1}$, Leandro M. Malard$^{7}$, Mauricio Terrones$^{4,2,6}$, Vincent H. Crespi$^{2,3,6,4}$, and Marcio D. Teodoro$^{1}$*}

\paragraph*{}

\noindent $^{1}$Departamento de F\'isica, Universidade Federal de São Carlos, São Carlos, São Paulo 13565-905, Brazil
\\
$^{2}$Department of Physics, The Pennsylvania State University, University Park, PA 16802, United States of America
\\
$^{3}$The Materials Research Institute, The Pennsylvania State University, University Park, PA 16802, United States of America
\\
$^{4}$Center for 2-Dimensional and Layered Materials, The Pennsylvania State University, University Park, PA 16802, United States of America
\\
$^{5}$Department of Electrical and Computer Engineering, University of Central Florida, Orlando, Florida 32816, USA
\\
$^{6}$Department of Materials Science and Engineering, The Pennsylvania State University, University Park, PA 16802, United States of America
\\
$^{7}$Departamento de F\'isica, Universidade Federal de Minas Gerais, Belo Horizonte, Minas Gerais 30123-970, Brazil
\\

\paragraph*{}

%
\newpage

\raggedright

\paragraph*{This Supporting Information includes:\newline}

\begin{itemize}
  \item Supporting Note 1: Power- and temperature-dependent photoluminescence for pristine WS$_2$ monolayer
  \item Supporting Note 2: Photoluminescence spectral analysis
  \item Supporting Note 3: Temperature-dependent photoluminescence for V-doped WS$_2$ monolayers
  \item Supporting Note 4: Additional magneto-photoluminescence data for $\sim$0.4$\%$ V-doped WS$_2$ monolayer
  \item Supporting Note 5: Magneto-photoluminescence data for $\sim$2.0$\%$ V-doped WS$_2$ monolayer
  \item Supporting Note 6: Sulfur vacancy changes the magnetic anisotropy energy in V-doped WS$_2$
\end{itemize}
\newpage


\section*{\large{Supporting Note 1: Power- and temperature-dependent photoluminescence for pristine WS$_2$ monolayer}}
\begin{figure}[htb]
	\centering
	\includegraphics[width=0.8\linewidth]{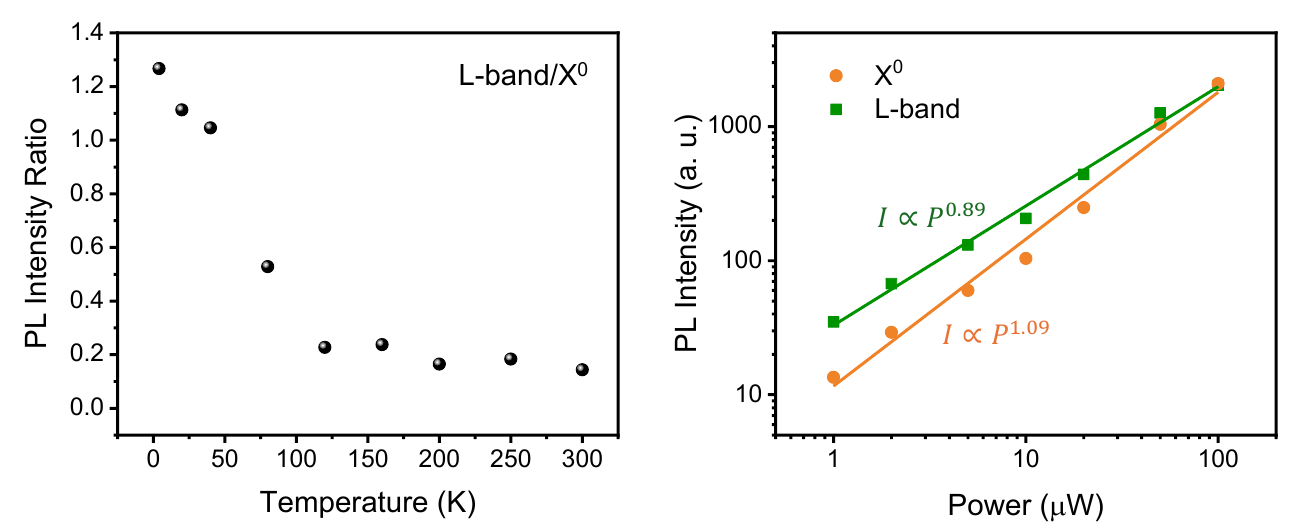} 
	\caption{\small 
    {\bf a} Temperature dependence for the L-band/X$_0$ PL intensity ratio for pristine WS$_2$ monolayer. A relative quenching in the L-band emission is observed above 100~K.
    {\bf b} L-band and X$_0$ PL intensity dependence on incident laser power. The L-band displays a sublinear power dependence, while X$_0$ shows a linear dependence. 
    }
	\label{SuppFig1}
\end{figure}

The measured PL spectra of pristine WS$_2$ monolayers exhibit free exciton (X$_0$) and defect-related L-band emissions. 
The L-band arises from radiative recombination involving localized mid-gap states introduced by defects. 
As the temperature increases, however, carriers trapped in these states can be thermally re-excited into delocalized band states and are increasingly subject to non-radiative recombination pathways, leading to a quenching of the defect-related emission.
Figure~\ref{SuppFig1}a highlights this effect, where the temperature-dependent L-band/X$_0$ PL intensity ratio quenches above 100~K.
Moreover, the localized character of these mid-gap states leads to a saturated occupation under increasing laser power.
This reflects in the sublinear power dependence of the L-band intensity shown in Figure~\ref{SuppFig1}b, whereas the X$_0$ state displays the expected linear dependence.


\newpage
\section*{\large{Supporting Note 2: Photoluminescence spectral analysis}}
\begin{figure}[htb]
	\centering
	\includegraphics[width=1.0\linewidth]{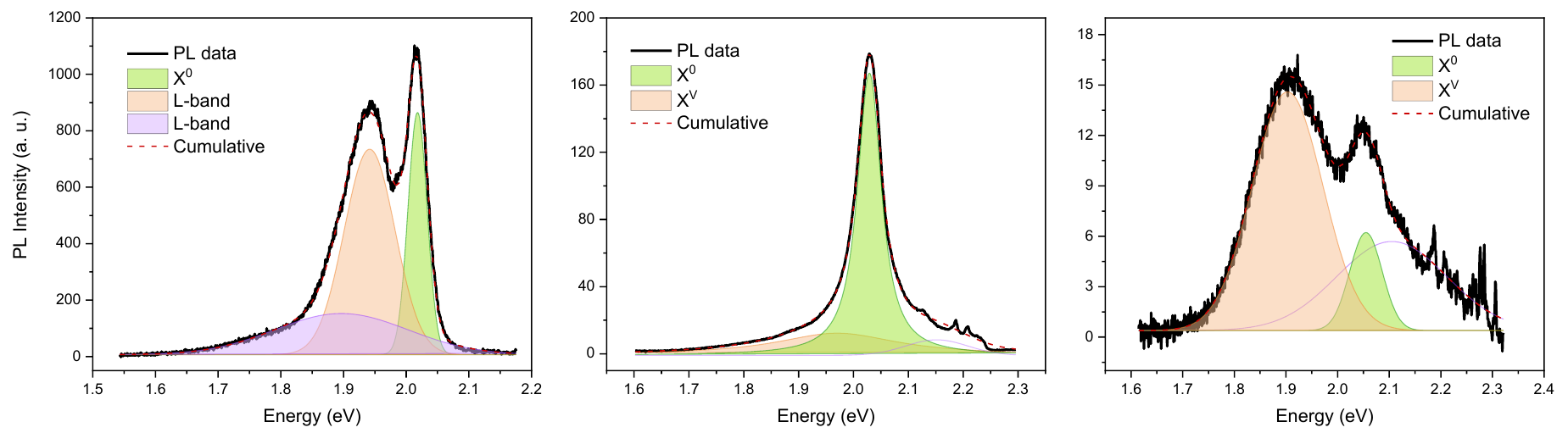} 
	\caption{\small 
    {\bf a-c} PL spectra analysis for pristine ({\bf a}), $\sim$0.4$\%$ ({\bf b}), and $\sim$2.0$\%$ ({\bf c}) V-doped WS$_2$ monolayers. The black solid lines represent the measured data, the dashed red lines show the cumulative fit, and the color-filled Gaussian peaks denote the fitted components for X$_0$, L-band, and X$_V$ emissions. For the doped samples, a higher energy peak (purple solid line) was also used to account for a laser background.
    }
	\label{SuppFig2}
\end{figure}

All PL spectra obtained in this study were fitted with Gaussian functions to analyze the integrated intensity and energy position parameters. 
For the pristine monolayers (Figure~\ref{SuppFig2}a), we fit the spectra with one peak for the X$_0$ emission (green) and two peaks for the L-band (orange and purple) due to its asymmetric character.
For the V-doped monolayers (Figures~\ref{SuppFig2}b,c), we fit the spectra with one peak for the X$_0$ emission (green), one peak for the X$_V$ emission (orange), and one auxiliary higher energy peak (purple) to account for a laser background --- which was negligible for the pristine sample due to its stronger signal.


\newpage
\section*{\large{Supporting Note 3: Temperature-dependent photoluminescence for V-doped WS$_2$ monolayers}}
\begin{figure}[htb]
	\centering
	\includegraphics[width=0.8\linewidth]{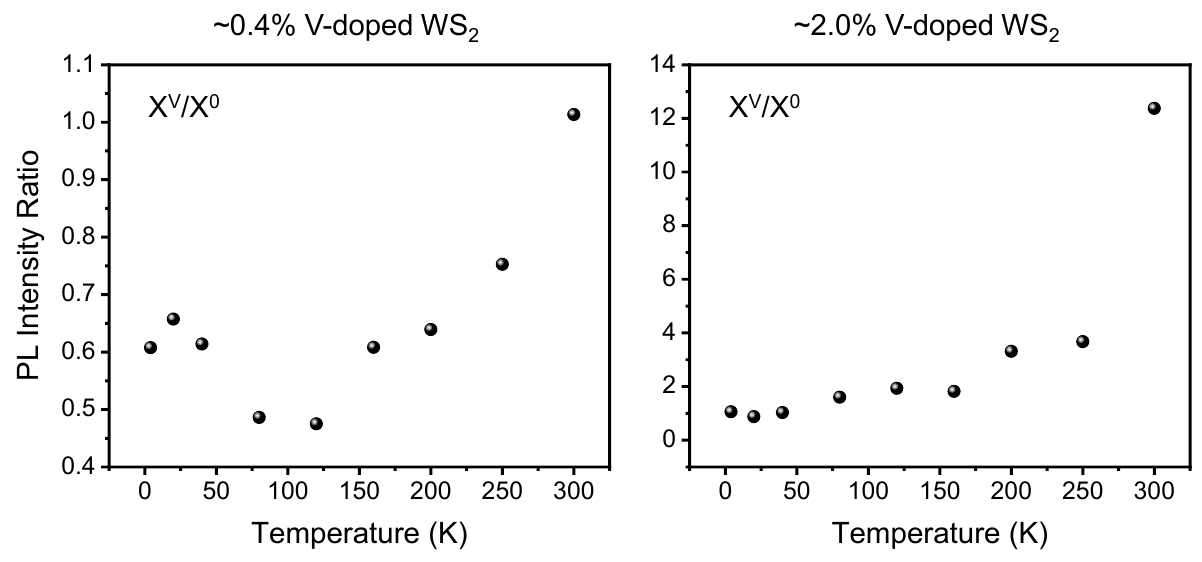} 
	\caption{\small {\bf a,b} Temperature dependence of the X$_V$/X$_0$ PL intensity ratio for $\sim$0.4$\%$ ({\bf a}) and $\sim$2.0$\%$ ({\bf b}) V-doped WS$_2$ monolayers. For both samples, the X$_V$ emission exhibits a finite signal for the entire temperature range and a relatively stronger intensity under increasing temperature.}
	\label{SuppFig3}
\end{figure}

The V-doped WS$_2$ monolayers display free exciton (X$_0$) and vanadium-related (X$_V$) emissions. In contrast with the L-band from pristine samples, the X$_V$ emission exhibits a finite signal from cryogenic up to room temperature.
This is explained by the non-localized character of V-like states, which introduce mid-gap bands instead of localized levels.
This response is evidenced in Figure~\ref{SuppFig3}, which shows a relatively stronger X$_V$/X$_0$ PL intensity ratio under increasing temperature.


\newpage
\section*{\large{Supporting Note 4: Additional magneto-photoluminescence data for $\sim$0.4$\%$ V-doped WS$_2$ monolayer}}
\begin{figure}[htb]
	\centering
	\includegraphics[width=1.0\linewidth]{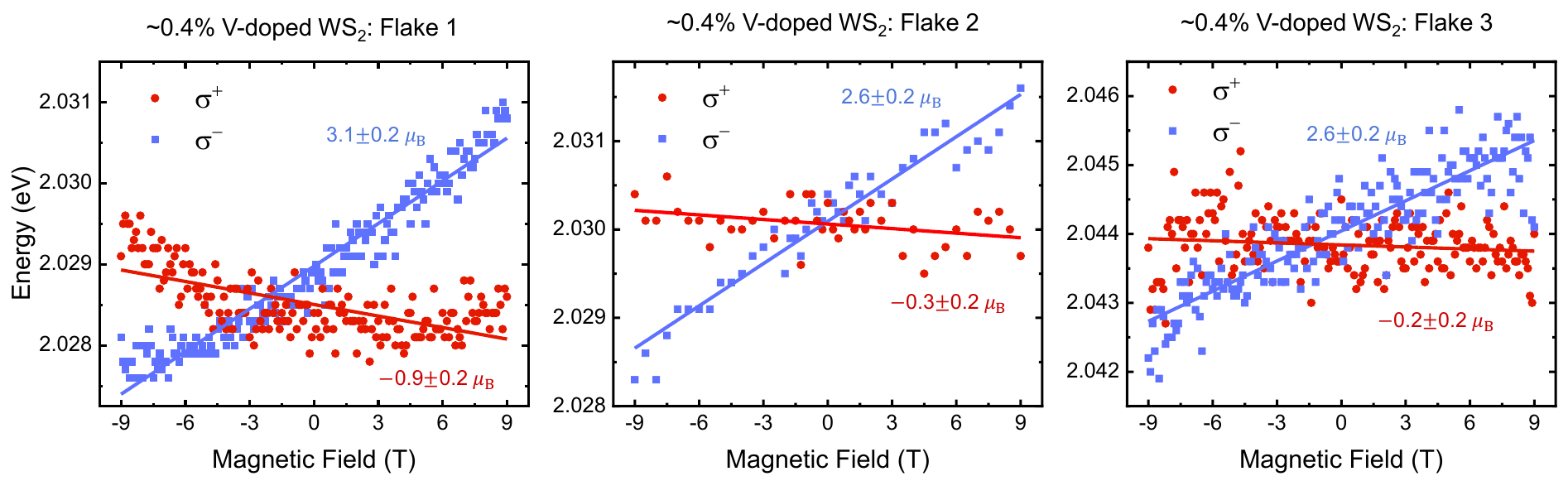} 
	\caption{\small {\bf a-c} Magnetic-field-dependent X$_0$ energy graphs for both $\sigma^+$ and $\sigma^-$ signals acquired for three distinct $\sim$0.4$\%$ V-doped WS$_2$ monolayers.
    The asymmetric Zeeman slopes are denoted in red and blue for the K and $-$K valleys, respectively.}
	\label{SuppFig4}
\end{figure}

To probe the reproducibility of the valley-asymmetric Zeeman shift for the $\sim$0.4$\%$ V-doped WS$_2$ monolayer, we carried out additional magneto-PL experiments in three different flakes.
The measurements and spectral analysis were performed similarly to those presented in the main text.
Figure~\ref{SuppFig4} shows the magnetic-field-dependent X$_0$ energies obtained at $\sigma^+$ and $\sigma^-$ detection.
Notably, all flakes corroborate the asymmetric Zeeman response, in which the $-$K valley Zeeman slope (\qty{2.6\pm0.2}{\mu_{\textnormal{B}}} $< \alpha_{-K} <$ \qty{3.1\pm0.2}{\mu_{\textnormal{B}}}) is significantly larger in modulus than the K valley Zeeman slope (\qty{-0.9\pm0.2}{\mu_{\textnormal{B}}} $< \alpha_{K} <$ \qty{-0.2\pm0.2}{\mu_{\textnormal{B}}}).
It is worth mentioning that the results shown in Figure~\ref{SuppFig4}b were acquired with steps of 0.5~T instead of 0.1~T.


\newpage
\section*{\large{Supporting Note 5: Magneto-photoluminescence data for $\sim$2.0$\%$ V-doped WS$_2$ monolayer}}
\begin{figure}[htb]
	\centering
	\includegraphics[width=0.8\linewidth]{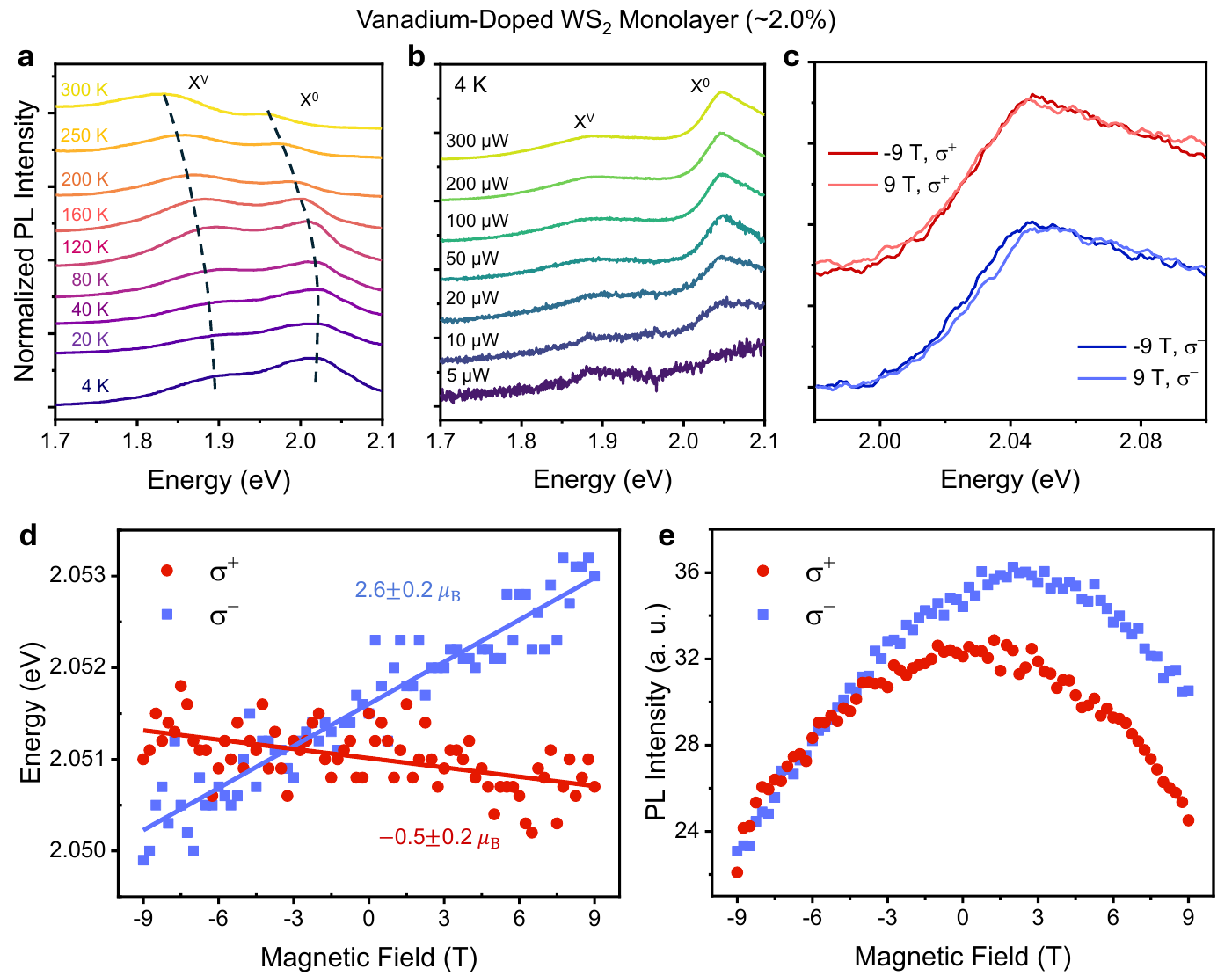} 
	\caption{\small {\bf a} Temperature- and {\bf b} power-dependent normalized PL spectra for a $\sim$2.0$\%$ V-doped WS$_2$ monolayer displaying X$_0$ and X$_V$ emissions. 
    {\bf c} Normalized PL spectra acquired at 9~T and $-$9~T, 4~K, linear excitation, and circular polarization detection for the X$_0$ emission in the $\sim$2.0$\%$ V-doped WS$_2$ monolayer.
    Red (blue) line spectra denote $\sigma^+$ ($\sigma^-$) detection.
    {\bf d} Magnetic-field-dependent X$_0$ energy graph for both $\sigma^+$ and $\sigma^-$ signals.
    Valley-dependent Zeeman slopes of $\alpha_K=$ \qty{-0.5\pm0.2}{\mu_{\textnormal{B}}} and $\alpha_{-K}=$ \qty{2.6\pm0.2}{\mu_{\textnormal{B}}} are obtained for K and $-$K valleys, respectively.
    {\bf e} Magnetic-field-dependent X$_0$ intensity graph for both $\sigma^+$ and $\sigma^-$ signals corroborating the valley asymmetric response.}
	\label{SuppFig5}
\end{figure}

To investigate how vanadium concentration impacts the asymmetric valley Zeeman effect of doped WS$_2$ monolayers, we also performed PL and magneto-PL experiments for the $\sim$2.0$\%$ doped samples. 
Figures~\ref{SuppFig5}a,b show temperature- and power-dependent PL spectra for the $\sim$2.0$\%$ V-doped WS$_2$ monolayer.
Similar to the $\sim$0.4$\%$ doped sample, we observe free exciton (X$_0$) and vanadium-related defect-bound exciton (X$_V$) emissions.
As evidenced in Figure~\ref{SuppFig3}b, the X$_V$ emission persists up to room temperature and with a relatively stronger intensity due to the higher vanadium concentration.
Figure~\ref{SuppFig5}c shows PL spectra (normalized by the maximum intensity) for the X$_0$ emission in the $\sim$2.0$\%$ V-doped WS$_2$ monolayer acquired at 9~T and $-$9~T, 4~K, linear excitation, and for $\sigma^+$ and $\sigma^-$ detections.
Despite the signal noise, a larger Zeeman shift at the $-$K valley than at the K valley is observed.
This asymmetry is better noted in Figure~\ref{SuppFig5}d, where the X$_0$ energies extracted by fitting the spectra with a Gaussian function are plotted for all measured magnetic fields. 
Through a linear regression of the scattered data, we obtained Zeeman slopes of $\alpha_K=$ \qty{-0.5\pm0.2}{\mu_{\textnormal{B}}} and $\alpha_{-K}=$ \qty{2.6\pm0.2}{\mu_{\textnormal{B}}} for K and $-$K valleys, respectively.
The modulation of the X$_0$ integrated intensity by the external magnetic field is also valley asymmetric, as shown in Figure~\ref{SuppFig5}e, with $-K$ valley presenting a greater overall intensity.

\newpage
\section*{\large{Supporting Note 6: Sulfur vacancy changes the magnetic anisotropy energy in V-doped WS$_2$}}
Introducing the magnetic moment into WS$_2$ cannot guarantee the breaking of the valley degeneracy---the direction is also important.
For example, if the magnetic moment on V is in-plane, we have two high-symmetry directions where the easy axis would typically align to.
Along the $x$-direction, the two valleys are still symmetry-related by the mirror $m_x$ operation; and if along the $y$-direction, they are symmetry-related by the \qty{180}{\degree} rotation along the $y$-direction.
Therefore, the in-plane dopant magnetic moment should not break the valley symmetry.
If the moment is out-of-plane, we will see the valley degeneracy breaking as shown in Fig.~4 in the main text.
However, for a vacancy-free freestanding V-doped WS$_2$, the easy axis is in-plane as shown in the first column of Tab.~\ref{tab:S1}.
\begin{table}[htb]
  \caption{The magnetic anisotropy energy (meV) for different configurations of sulfur vacancy (S-vac) in our computational model of V-doped WS$_2$.
    V/W/S atoms are colored as red/gray/yellow, respectively.
    The dashed circle shows the position of the S-vac in the model.
    Values are referenced to the lowest one in each column.
    The easy axis is in-plane when there is no S-vac or the vacancy is near the V-dopant.
    The easy axis can be out-of-plane when the S-vac is not near the dopant.
    \label{tab:S1}}
  \begin{tabular}{b{1.8cm}ccc}
    \toprule
    magnetic moment direction & \includegraphics[width=0.25\textwidth]{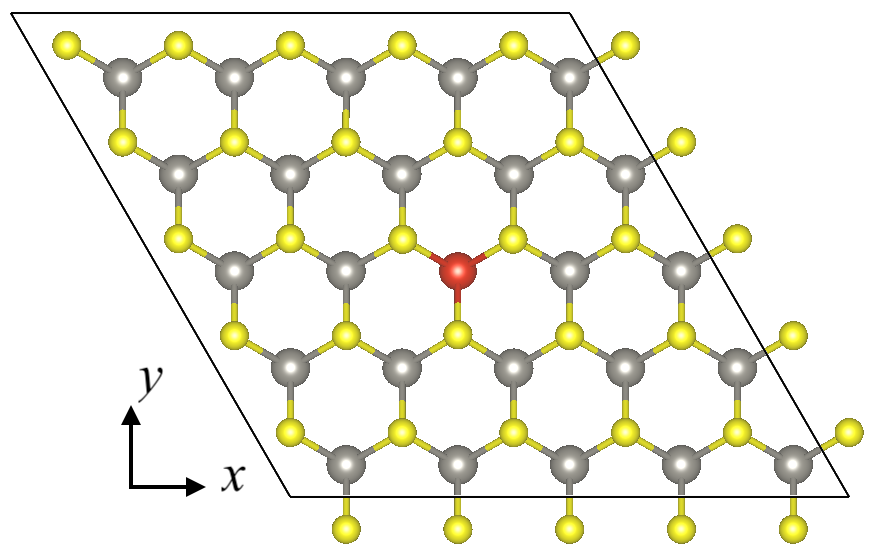} & \includegraphics[width=0.25\textwidth]{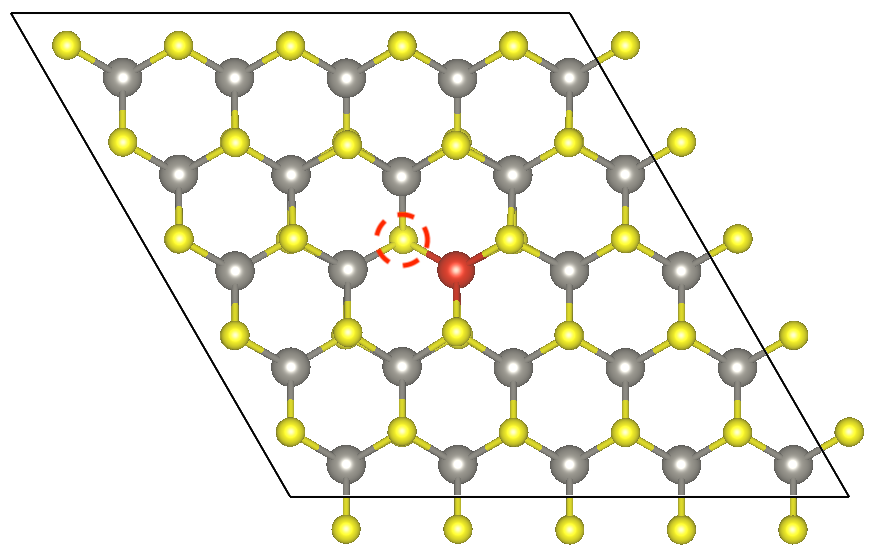} & \includegraphics[width=0.25\textwidth]{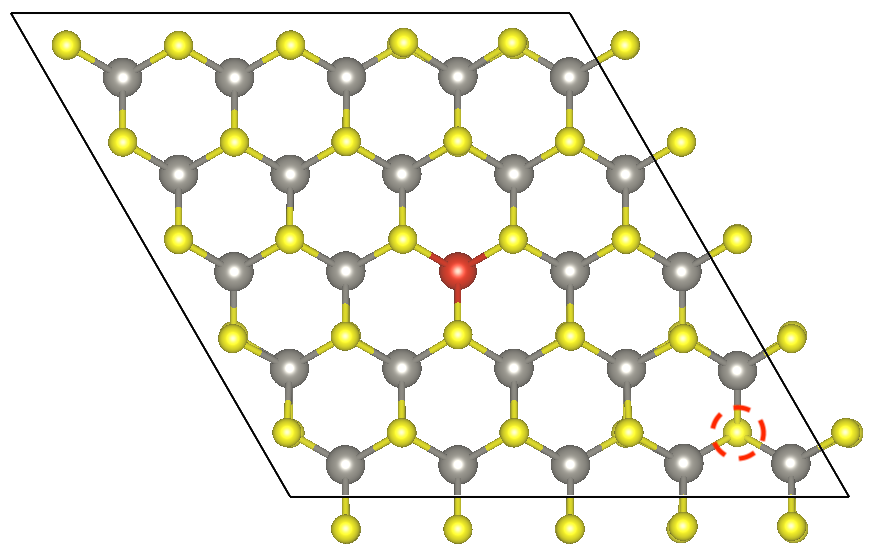} \\
     & no S-vac & S-vac near the V & S-vac not near the V\\
    \midrule
    $x$ & 0.00 & 0.02 & 0.11 \\
    $y$ & 0.00 & 0.00 & 0.11 \\
    $z$ & 0.07 & 1.08 & 0.00 \\
    \bottomrule
  \end{tabular}
\end{table}
From the magnetic anisotropy energy calculation (Tab.~\ref{tab:S1}), we notice that sulfur vacancies that are a couple of lattice constants away from the V-dopant could bring the easy axis out-of-plane.
Although the cases where the sulfur vacancy not right next to the dopant have higher total energies, these configurations can happen in more locations thus having higher configurational entropy, and we indeed see a significant amount of these vacancies experimentally \cite{zhang2020monolayer}.
Together with the possible effects from the substrate, we expect it possible to have domains with the easy axis out-of-plane.

\providecommand{\latin}[1]{#1}
\makeatletter
\providecommand{\doi}
  {\begingroup\let\do\@makeother\dospecials
  \catcode`\{=1 \catcode`\}=2 \doi@aux}
\providecommand{\doi@aux}[1]{\endgroup\texttt{#1}}
\makeatother
\providecommand*\mcitethebibliography{\thebibliography}
\csname @ifundefined\endcsname{endmcitethebibliography}  {\let\endmcitethebibliography\endthebibliography}{}

\end{document}